\documentstyle[12pt]{article}
\begin{document}

\title{Theory of Electromagnetic Wave Transmission through Metallic Gratings of Subwavelength Slits.}
\author{N.Garcia$^{1}$ and  M.Nieto-Vesperinas$^{2}$
 \and $^{1}$Laboratorio de F\'{i}sica de Sistemas Peque\~{n}os y
Nanotecnolog\'{i}a, \and Consejo Superior de Investigaciones
Cientificas , \and Serrano 144, 28006 Madrid, Spain. \and $^{2}$
Instituto de Ciencia de Materiales de Madrid, \and Consejo Superior
de Investigaciones Cient\'{i}ficas, \and Campus de Cantoblanco
Madrid, Spain. \and nicolas.garcia@fsp.csic.es;  mnieto@icmm.csic.es
}
%EndAName}
%\date{\today}
\maketitle
\begin{abstract}
We present FDTD calculations for transmission  of light and other
electromagnetic waves through periodic arrays of slits in a metallic
slab. The results show resonant, frequency dependent, transmittance
peaks for subwavelength widths of the slits which can be up to a
factor of ten with respect to those out of resonance. Although our
conclusions agree with previous work by Lezec and Thio as regards
both the magnitude of the enhancement and the lack of contribution
of surface plasmon polaritons of the metal surface to this effect,
we derive an interpretation from a theory that deals with emerging
beam-Rayleigh anomalies of the grating, and with Fabry-Perot
resonances of the perforated slab considered as an effective medium.
\end{abstract}

\section{Introduction}
 Extraordinary optical transmission through hole
arrays of subwavelength diameter in metal films has been subjected
to extensive study since it was first reported the detection of
transmitted intensity large enhancements, which were polarization
dependent \cite{Ebbesen} and attributed to excitation of surface
plasmon polaritons (SPPs) of the film surface \cite{Porto},
\cite{G-Vidal-Lezec}. Nevertheless, recently both the enhancements
of the transmission and the role of the SPPs in the phenomenon have
been reconsidered \cite{Thio1}, \cite{Thio2} showing that neither
the enhancements are as large as initially proposed, nor the SPPs
play a key role since the same observations are obtained in films of
materials like Cr which has a very large permittivity imaginary part
thus defining a very dissipative and broad plasmon. The same occurs
in W and Si films that have a positive real part of the dielectric
permittivity in the visible, and which of course do not support
SPPs. These experiments arose a certain controversy as regards the
understanding of the underlying physics behind phenomenon, and have
thus warranted a closer look at the interpretation of these
transmittance enhancements as in a recent publication \cite{Bai},
where it is shown that the initially reported enhancements of
several orders of magnitude are so when the ratio is taken with
respect to the transmitted intensity obtained from Bethe's theory
\cite{Bethe}. However, Bethe's approximation, only valid for
apertures much smaller than the wavelength and that thus assumes the
field  as constant inside them, is not applicable to this
configuration and thus the aforementioned ratios are misleading.
More realistic is to take the ratio of the resonantly enhanced
intensities with respect to those obtained, both in experiments and
exact calculations, out of resonance as done in \cite{Thio1} and
\cite{Bai}. Then, these enhancements are larger than without
resonant transmission not by orders of magnitude  but by a factor
that in data obtained so far is not larger than 10.

In this paper we present an interpretation of the enhanced
transmission of light through a periodic array of slits in a metal
slab by means of a model that combine an approximate analytical
calculation of scattering from corrugated metallic interfaces with
that of transmission through metal slabs. The results agree
qualitatively with those of \cite{Thio1} and \cite{Bai} for circular
holes. We conclude that transmission peaks are produced either by
those kind of Wood anomalies known as Rayleigh anomalies (RA),
namely those due to the emerging beams appearing in the passing-off
of a new diffracted order, \cite{Fano}, \cite{Hessel-Oliner},
\cite{Garcia}, \cite{MNV}, and that are well known to giving rise to
strong discontinuities of the diffracted intensities, or as a result
of the Fabry-Perot (FP) type resonance in the film layer as if it
were homogeneous with an effective refractive index. The FP
resonance comes from reflection and transmission at the entrance and
the exit interfaces of the slab. We shall show results both by an
exact FDTD calculation and by means of a theoretical model that
shows the physical meaning and nature of the enhancement phenomenon
in the slit grating transmittance. In this connection, it should be
pointed out that there is agreement that this kind of resonances
appear in these slit arrays as recently shown in  \cite{Park1},
\cite{Park2}, \cite{Catrysse}; also related are  those so-called
horizontal and vertical surface plasmon resonances by the authors of
ref.\cite{Collin}. Other researchers like to interpret them in terms
of coupling of SPPs at both surfaces of the film and of cavity modes
in the slits \cite{Moreno}. Further studies along these lines can
also be found in \cite{Popov}. In addition, the interpretation of
waveguide mode resonances and diffraction as responsible for the
extraordinary transmission is pointed out in \cite{Lalanne} and
\cite{Cao} where it is observed that transmittance of subwavelength
metallic gratings may be nearly zero for frequencies corresponding
to SPP excitation. Also, the introduction of film modulations to
obtain similar resonant effects can be found in \cite{Bonod}.

We shall first show how one slit does not transmit for s
polarization but it does so for p-polarization. Then we shall argue
that the transmission enhancement  is not due to the SPP of the
metal surface because with the slits present in the film this
polariton is too extended and thus only one of its many Fourier
components matches thus being not enough to yield the enhancement .
By contrast, the Rayleigh anomaly is by contrast very sharply
defined: that wavelength at which the z-component of the diffracted
beam wavevector vanishes.

As for the FDTD calculations \cite{Taflove}, we address grids with a
space step $\delta x=40nm$ and a time step $\delta t=0.1\delta x/c$,
where $c$ is the light velocity. This sampling yields good
convergent solutions. Periodic boundary conditions where used along
the boundary parallel to the main direction of light propagation. On
the other hand,  the uniaxial perfectly matched layer (UPML)
absorbing boundary condition, was used on the boundary perpendicular
to light propagation; (showing no significantly different results
for the particular structure here used to previous calculations done
by means of Liao's conditions). Concerning the materials in these
calculations, and as far as Ag is concerned, where SPPs have   long
been studied both in flat and periodically corrugated surfaces
\cite{Raether}, \cite{Garcia2}, \cite{Madrazo}, and for which the
imaginary part $\epsilon_2$ of its permittivity $\epsilon=
\epsilon_1+ i \epsilon_2$ is small enough so that the SPP is well
defined \cite{Raether} in Drude's model with a bulk plasmon
wavelength $\lambda_p=325nm$ $\omega_p\sim 3.8eV$, it is known  that
for example in a periodic  Ag interface with a small single Fourier
component, the enhancement of the field due to the SPP may reach
large values (100 times that of the incident field) \cite{Garcia2}.
However, when the amplitude of this harmonic  increases, or many
Fourier component exist, (this happens in e.g. a random rough
interface
 \cite{MNV} or in a step-like profile of the surface like in the slits addressed here), the enhancement is
 drastically reduced due to the enlargement of the SPP linewidth.
In our calculations, the metal is characterized in Drude's model
$\epsilon(\omega)/\epsilon_f=1-(\omega_p^2/\omega^2+2i\omega\delta))$
by fitting its permittivity $\epsilon_f$, the bulk plasma frequency
$\omega_p$ and the damping constant $\delta$ to the experimental
data by Johnson and Christy \cite{Christy} for Ag. The chosen
parameters are $\epsilon_f=6.8$, $\omega_p\sim 3.8eV$ and
$\delta\sim -0.02eV$.

\section{Calculations}
We consider a plane electromagnetic wave normally incident on a
lamellar grating consisting of 1-D slits as indicated in Fig.1,
which we shall characterize by a profile function $z=D(x)$ with
respect to the entrance plane of the film: $z=0$. This configuration
was studied in \cite{Porto}. The parameters are: period $d=3.5 \mu
m$, slit width $a=0.5\mu m$; the thickness of the film $h$ varies
from $0.2\mu m$ to $4\mu m$.

We shall write the intensity  transmitted by the grating in the far
zone as:

\begin{equation}
I(Q_m)=\frac{q}{q_0}|A(Q_m)|^2,
\end{equation}

where the incident and diffracted wavevectos are: ${\bf
k_0}=(K_0,q_0)$, ${\bf k_m}=(Q_m,q_m)$,
$k_{m}^{2}=k_0^2=(2\pi/\lambda)^2$.   $Q_m=K_0+2\pi m/d$,
$m=0,1,2,...$, $q_m=\sqrt{k_0^2-Q_{m}^{2}}$.

The diffracted amplitude transmitted by the grating is written as
\cite{Garcia-MNV}:

\begin{equation}
A(Q_m)=\frac{1}{q_m}\int_{0}^{d}dx'F(x')\exp[-i(Q_m x'+q_m D(x'))].
\end{equation}

In Eq.(2) the source function $F(x)$ is given by the boundary
conditions at the grating surface. On the other hand, for normal
incidence: $Q_m=2\pi m/d$.

In order to explicitely compare the contribution  of the FP
resonance as compared with that due to the RA, we shall write $F(x)$
as:

\begin{equation}
F(x)=\exp[iq_0 D(x)]\sqrt{T}\exp(i\phi),
\end{equation}

$\phi$ being a certain phase and $T$ being the film trasmittance:

\begin{equation}
T=\frac{1}{(1-r)^2} \frac{1}{1+4r/((1-r)^2 \sin^2(\delta/2))},
\end{equation}

where $\delta=q_0 n_{eff}h+2\psi$, $n_{eff}$ is an effective index
of the film averaged from the opaque and transparent portions of the
slab, which makes sense for small wavelengths, and $\psi$ denotes a
phase shift between the incoming and the reflected wave.

In  this way, the diffracted intensity for a given $Q_m$ becomes:

\begin{equation}
I(Q_m)=T |B(Q_m)|^2.
\end{equation}

$B(Q_m)$ is obtained from Eqs.(2) and (3). We then perform the
integration for the profile $D(x)$ corresponding to the slit
grating. The result is:

\begin{equation}
B(Q_m)=4i\frac{q_0}{q_m}\sin(\frac{\pi m
a}{d})\sin[\frac{(q_0-q_m)h}{2}].
\end{equation}

The transmittance is next calculated by integrating $I(Q_m)$ over
all diffraction orders, namely, at all angles of transmission.
Figs.2(a)-2(d) correspond to the case dealt with in \cite{Porto}.
Both FDTD calculations (solid line) and results from Eq.(5) (broken
line) are plotted. The slits are considered as practiced in an Ag
slab of thickness $h$ which varies from $0.2 \mu m$ to $4 \mu m$,
the period being $d=3.5 \mu m$. Of course, the agreement between
calculations and the analytical theory of Eqs. (2)-(6) is
qualitative: these equations help us to interpret the origin of the
resonant peaks of the transmission intensity, even though the
numerical value of their results does not coincide with the exact
one provided by the FDTD calculation.

 We observe peaks under p-polarization, (namely with the $H$
vector along the $y$-axis in Fig.1). In this case there is no
cut-off for transmission because the first mode of the aperture
cavity has the index (1,0); namely, it corresponds to a homogeneous
mode. This is the underlying reason why there is larger transmission
for p-polartization. Conversely, under s-polarization,  there is a
frequency cut-off and thus an evanescent transmitted mode through
the aperture cavity which has the index (1,1), and hence there is no
transmitted intensity since this cut-off exponentially dampens its
intensity. (An illuminating discussion of this effect is presented
in Jackson's text \cite{Jackson}, Section 8.4). Only by overcoming
this cut-off at high frequencies a transmitted intensity arises in
s-polarization.

On the other hand, in these figures, which show agreement with the
calculation of \cite{Porto}, the two peaks, (in particular in Fig.
2(a) corresponding to $h=3\mu m$ and $n_{eff}=1.2$), are well
separated from each other. Neither of these peaks arise from surface
plasmon polaritons (SPPs). The broader peak in the right is a FP
resonance due to the contribution of $T$ in Eq.(5). The narrower
peak, placed at shorter wavelengths in Fig. 2(a), is a Rayleigh
anomaly, namely one due to the passing-off of a new diffracted mode;
as such it occurs at a sharply well defined value of the wavelength,
namely that at which $q_m=0$. As the slab becomes thinner, the FP
resonance peak shifts to smaller wavelengths and tends to overlap
with the Rayleigh-emerging beam resonance. This is shown in Figs.
2(b)-2(d) corresponding to $h=2 \mu m$, $h=1.2 \mu m$ and $h=0.6 \mu
m$. In fact, Fig.2(d) shows that in the latter case the peak
resonance is mainly from the pure emerging beam. In the cases of
Figs.2b) and 2(c), there is an overlapping of the FP and the
emerging beam resonances with a subsequent splitting of both peaks.
Therefore, this combination of the FP plus the emerging beam
resonances explain all the structure of the frequency dependent
transmittance curves given by the exact FDTD calculations, in
numerical agreement with those of \cite{Porto}.

It should be pointed out that, although the resonance frequency of
SPPs of the flat Ag slab is very close to that of the RA of this
grating, the large corrugation introduced on its surface by the
presence of the slits, in addition to to its broad Fourier spectrum
as pointed out before, broadens the scattered field angular spectrum
which, consequently, only weakly interacts with the SPP of the film
flat surface.  This is well known from the theory of surface-relief
grating diffraction \cite{Garcia2}, \cite{MNV}, \cite{Madrazo},
where it is shown how the absorption of light by a grating, due to
the SPP excitation, decreases as its depth increases. In addition,
as we shall see later, these RA peaks equally appear for a non metal
slab with slits as Cr.

As the slab thickness $h$ increases, more FP resonances appear at
longer wavelengths. This is shown in Fig.3(a)  for $h=4\mu m$
which corresponds to the calculation using Eq.(5) and that now
exhibits two of these FP peaks: one near $5000 nm$ and an
additional broader one at $10000 nm$. As remarked above, none of
these peak structures appears in s-polarization; this is seen in
Fig.3(b) by the FDTD results. In this last figure we show the
transmission curve slowly rising from the wavelength of $1.5 \mu
m$ and more abruptly towards the left to $1 \mu m$, which is
precisely double of the slit width $a=0.5\mu m$, and thus
constitutes the cut-off wavelength. This further demonstrates that
the observation of transmission in these structures under
p-polarization only, is not caused by SPP resonances, but is
rather due to the above mentioned transmission characteristics of
the waveguide that constitutes the aperture. We would also like to
stress that these transmission enhancements, obtained at
resonance, are not orders of magnitude larger, but given by
factors not larger than 10.

The main contribution of both FP and Rayleigh anomalies to this
transmission behavior, and not of SPPs, is further shown in Fig.3(c)
which corresponds to a slit grating perforated in a Cr slab of
thickness $h=0.6\mu m$. Namely, identical to that dealt with in
Fig.2(d) except for the material that now has a completely different
dielectric constant. In fact, its imaginary part that defines the
sharpness of the SPP at the frequencies of Fig.3(c) is 10 times
larger than for Ag. The transmitted intensity exhibits not only the
same RA resonance, but also an almost identical transmission curve
as that of the Ag grating at all frequencies, as one sees on
comparison of Figs. 2(d) and Fig.3(c). This result agrees with
previous reports stating that these transmission peaks  are equally
obtained in slits practiced in metals and in dielectrics
\cite{Thio1}, \cite{Thio2}, \cite{Thio3}, \cite{Sarrazin}.

On the other hand, as mentioned in Section 1, SPPs are well defined
in Ag and Au in the visible . Then the width of the plasmon is small
and its lineshape is sharp,(see \cite{Raether}) . However, at the
frequencies at hand the values of the imaginary part of Ag and Cr
are about 5-35 and 44-200, respectively. This produces, (even more
for Cr), a quite broad, and hence badly defined, SPP, if any. In
addition, a SPP is a well defined harmonic wave. This is so when the
roughness of the suface, (i.e. the $h$ parameter in our grating)
that holds the plasmon, is low. This leads to large enhancements in
the near field intensity. However as the roughness increases, the
plasmon width increases, and then the field enhancement is
drastically reduced \cite{Garcia2}, \cite{MNV}. Hence, SPPs cannot
significantly contribute to these transmission peaks. On the
contrary, for the Rayleigh resonances obtained in these
calculations, the trasmitted intensity increases with  the roughness
$h$ up to a certain value, like when the corrugation is a harmonic
function \cite{Garcia}, \cite{MNV}.

\section{Conclusions}
We have put forward an analytical approximate model, validated by
FDTD calculations, that accounts for the transmitted intensity
enhancements of light and other electromagnetic waves transmitted
through a grating consisting of slits in a slab. The model has an
interpretative value as it shows that these enhancements are due to
both Rayleigh anomalies of the grating and to Fabry-Perot resonances
of the slab. The contribution of surface plasmon polaritons to this
effect is negligible, although when the apertures have other
geometry, like circular holes, their morphological resonances
(localized plasmons, not to confuse with SPPs) do also play an
important role in these enhancements \cite{Bai}.  On the other hand,
the enhancements obtained are a factor not larger than ten, but not
by orders of magnitude, larger than out of resonance. We notice that
there are critical experimental data that observed similar
enhancement factors for Ag, W and Si in gratings of holes
\cite{Thio1}, \cite{Thio2} in the visible, where the last two
materials are dielectric, (namely, they have a positive real part of
the dielectric constant). In this case, it is well know that no SPP
resonances can exist.

The results of this paper support those of  Ref.4, although the
analysis there, based on the angular spectrum of plane waves for the
scattered field above and near apertures,  attributes the
enhancements to the contribution of a broad bundle of evanescent
waves of this angular spectrum. As far as we know, that theory,
which is also scalar, has not been tested with any exact procedure.
We agree with the experiments and magnitude of the enhancements
obtained in \cite{Thio1}, as well as with the lack of contribution
of SPPs to these peaks, as discussed there. We believe that our
analytical calculations, validated by the FDTD simulations , help to
support this agreement, even though it is not our purpose to enter
in a direct comparison of our theory with that of \cite{Thio1}.  We
believe, nevertheless, that the mechanisms discussed here are
sufficient to understand these enhancements as far as slits are
concerned, in the same way as a recent publication done with the
same procedure (cf. \cite{Bai}) explains the enhancements for arrays
of holes. In both cases: holes and slits, the characteristics of
transmission of just one of these apertures determines the resultant
properties of the array. However, whereas the properties of the
transmitted modes in each slit for either s or p-polarization rules
the enhancements of the array as discussed in Section 2; in the case
of circular holes, the morphological resonances of each hole are a
crucial factor that contributes to the transmission peaks in
addition to that of the Rayleigh anomalies of the arrays.

We hope that the arguments presented in this work on the physics of
this effect will shed light on discussions  as regards the nature of
these enhancements. \vspace{1cm}

This work has been supported by the DGICYT and European Union. We
thank K. Ponizovskaya for assistance with the computer calculations.

%\bigskip

\newpage

{\bf Figure Captions}

\begin{itemize}

\item{Figure 1: Scattering geometry. Period $d=3.5 \mu
m$, slit width $a=0.5\mu m$; the thickness of the slab $h$ varies
between $0.2\mu m$ and $4\mu m$.}

\item{Figure 2: p-polarization. Transmittance versus illumination
wavelength for the Ag grating parameters: period $d=3.5 \mu m$, slit
width $a=0.5\mu m$. Solid line: FDTD calculation. Broken line:
Theory calculation based on Eqs.(4)-(6) (a) Thickness $h=3\mu m$;
($n_{eff}=1.2$ in theory calculation).(b) Thickness $h=2\mu m$;
($n_{eff}=1.4$ in theory calculation). (c) Thickness $h=1.2\mu m$;
($n_{eff}=1.7$ in theory calculation). (d) Thickness $h=0.6\mu m$;
($n_{eff}=3.1$ in theory calculation).}

\item{Figure 3: (a) p-polarization. Transmittance versus illumination
wavelength for the Ag grating parameters: period $d=3.5 \mu m$, slit
width $a=0.5\mu m$, thickness $h=5\mu m$. Calculation based on
Eqs.(4)-(6) with $n_{eff}=1.2$. (b) s-polarization. Transmittance
versus illumination wavelength for the Ag grating parameters: period
$d=3.5 \mu m$, slit width $a=0.5\mu m$, thickness $h=0.6\mu m$. (c)
p-polarization. FDTD calculation for the same grating as in
Fig.2(d), but now the material is Cr.}

\end{itemize}

\end{document}